\documentclass[sigconf]{acmart}
\usepackage{dblfloatfix}  
\usepackage{graphicx}
\usepackage{color}
\usepackage{hyperref}
\usepackage{tabularx}
\usepackage{makecell}
\setcopyright{none}
\renewcommand\footnotetextcopyrightpermission[1]{} 
\usepackage{graphicx} 
\usepackage{tabularx}
\usepackage{hyperref}
\usepackage{placeins}
\usepackage{enumitem}
\usepackage{titlesec}
\usepackage{booktabs} 
\usepackage{array} 

\titlespacing{\subsection}{0pt}{*0.8}{*0.5}
\titlespacing{\subsubsection}{0pt}{*0.5}{*0.5}

\newcolumntype{Y}{>{\centering\arraybackslash}X}
\newcolumntype{Z}{>{\raggedright\arraybackslash}X}

\AtBeginDocument{%
  \providecommand\BibTeX{{%
    \normalfont B\kern-0.5em{\scshape i\kern-0.25em b}\kern-0.8em\TeX}}}

\acmConference[arxiv]{}{2023}{}

\begin{document}

\title{Eliciting Topic Hierarchies from Large Language Models}

\author{Grace Li}
\email{gl2676@columbia.edu}
\affiliation{%
  \institution{Columbia University}
  \city{New York}
  \state{New York}
  \country{USA}
}

\author{Tao Long}
\email{long@cs.columbia.edu}
\affiliation{%
  \institution{Columbia University}
  \city{New York}
  \state{New York}
  \country{USA}
}

\author{Lydia B. Chilton}
\email{chilton@cs.columbia.edu}
\affiliation{%
  \institution{Columbia University}
  \city{New York}
  \state{New York}
  \country{USA}
}

\renewcommand{\shortauthors}{Li et al.}

\begin{abstract}
Current research has explored how Generative AI can support the brainstorming process for content creators, but a gap remains in exploring support-tools for the pre-writing process. Specifically, our research is focused on supporting users in finding topics at the right level of specificity for their audience. This process is called topic scoping. Topic scoping is a cognitively demanding task, requiring users to actively recall subtopics in a given domain. This manual approach also reduces the diversity of subtopics that a user is able to explore. We propose using Large Language Models (LLMs) to support the process of topic scoping by iteratively generating subtopics at increasing levels of specificity: dynamically creating topic hierarchies.  We tested three different prompting strategies and found that increasing the amount of context included in the prompt improves subtopic generation by 20 percentage points. Finally, we discuss applications of this research in education, content creation, and product management. 
\end{abstract}




\settopmatter{printfolios=true}
\maketitle

\section{Introduction}
Generative AI has changed the way that content creators brainstorm ideas and structure content \cite{wang2024reelframer}, \cite{Tweetorial_Hook}. But using models like ChatGPT to directly generate content often results in bland and inauthentic results \cite{calderwood2020novelists}. As a result, many have instead focused on leveraging the ability of Generative AI in the formative stages of content creation \cite{shaer2024ai}. Recent work has explored the effectiveness of AI in the brainstorming process, but not much research has been done to explore the prewriting process \cite{Tweetorial_Hook}. Before a writer begins the process of outlining, they must determine a topic of the right scope to fit the venue and medium that they will publishing their work on. They must consider the constraints such as character-count when publishing on Twitter and the length of their video when publishing on TikTok or Instagram reels. This process of finding a topic at the appropriate level of specificity is called topic scoping. Topic scoping is used by educators when creating lesson plans, journalists when finding angles on current events, and researchers when determining specific projects to pursue as part of a grant.

The main challenge of topic scoping is iteratively breaking a broad domain, like Computer Science, into smaller and smaller subtopics to help the user pick a more specific subject area. For example, when picking a topic to teach within User Interface Design, a teacher might decompose UI into Usability Heuristics, and then break it down further into Visual Information Design. Through the process of iterative topic scoping, we can generate topic hierarchies--information trees where each node is a subtopic under the root and each level of the tree represents an increasing level of subtopic specificity. The Dewey Decimal System for book classication in libraries and the Taxonomy of Life that classify organisms by their Kingdom, Phyllum, Class etc. are examples of topic hierarchies that have been established and exist in the world. But managing the creation of topic hierarchies remains a non-trivial task. Currently, the number established topic hierarchies are limited and  static--the content often becoming outdated when they are not maintained. The maintenance of topic hierarchies is time and labor intensive task.


\begin{figure*}[ht]
    \centering
    \includegraphics[width=1\textwidth]{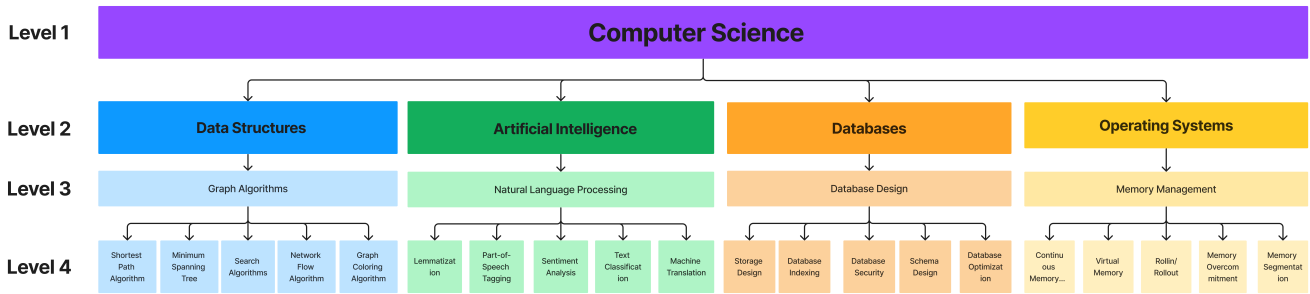}
    \caption{Topic hierarchy from Wikipedia's ``Category: Computer science'' with certain missing subtopics being filled in from college level curriculum. This Topic hierarchy is used as a test suite to evaluate the three different prompting strategies to generate subtopics in Level 2, Level 3, Level 4, and Level 5.}
    \label{fig:sample_category}
\end{figure*}

We explore how Large Language Models (LLMs) are able to dynamically elicit topic hierarchies to support the process of topic scoping. On their own, LLMs struggle with generating fine grain subtopics in more specific domains. Like users, as topics get more niche, machines struggle to generate subtopics that are unique, related, and specific to the given topic. We test three different prompting strategies on five different levels of topic specificity--with an emphasis on generating subtopics at the most specific level. The three conditions that we explored were 

\begin{enumerate}[noitemsep]
    \item Current Topic: ``List 5 subtopics in \textit{Natural Language Processing}.''
    \item Root + Current Topic: ``In \textit{Computer Science}, list 5 subtopics in \textit{Natural Language Processing}.''
    \item Full Path + Current Topic: ``In \textit{Computer Science} and \textit{Artificial Intelligence}, list 5 subtopics in \textit{Natural Language Processing} ''
\end{enumerate}

We used two annotators to evaluate the subtopics generated from the three different prompting techniques and measure the appropriateness of the generated subtopics on the relatedness, repetitiveness, and specificity. 

The Full Path + Current Topic prompting technique improved subtopic generation by 21 percentage points. This finding is inline with previous findings that show that in-context learning improves language model generations \cite{dong2022survey}. This paper demonstrates that LLMs can assist in the topic scoping process, helping create a structured exploration of possible subtopics within a given domain and proposes future work around designing an interactive system to support users in the process of topic scoping. 

\section{Related Work}
LLM-support in topic scoping and the creation of topic hierarchies extends research in the following domains: Approaches in Extracting and Structuring Information and LLMs on Various Knowledge Tasks. 


\subsection{Approaches in Extracting and Structuring Information}
Topic hierarchies help facilitate the process of topic scoping by structuring knowledge into a tree-like structure. Topic hierarchies fit into a broader domain of knowledge structuring. Like knowledge graphs that seek to unify information across disparate sources \cite{gutierrez2021knowledge}, to streamline the process of knowledge retrieval \cite{yan2018retrospective}, topic scoping seeks to add structure and organization to existing information. Furthermore, knowledge graphs have been used to support knowledge retrieval and brainstorming for different user centered systems \cite{ji2021survey}. Sun et. al. uses information propagation to add multimodal components to knowledge graphs for recommendation systems to support user-item interaction  \cite{sun2020multi}. Ait-Mlouk et. al.  used a support vector machine (SVM) for intent classification to query a knowledge base to support domain-specific question and answering \cite{ait2020kbot}. 

Broadly, the idea of topic scoping is closely related to text mining--transforming unstructured knowledge into a structured format to find patterns and to make new discoveries \cite{tan1999text}. Text mining has been explored within the context of education to support online learning \cite{ferreira2019text}. Mansur et. al. used natural language processing techniques like SVM and K-means for text classification to group educational resources together \cite{mansur2013social}. Similarly, Crossley et. al. incorporated student demographic information as features in their automatic essay evaluation system to train a regression model \cite{crossley2015pssst}. Jin et. al. used the  Blei's latent Dirichlet Allocation (LDA) to classify financial news articles and to obtain each article's topic distribution to train a linear regression model to make movement forecasts on the foreign currency market \cite{jin2013forex} \cite{blei2003latent}. 

Finally, the process of topic scoping is related to research in topic discovery and topic retrieval. Work in topic discovery includes characterizing the different topics within Twitter threads using a partially supervised learning model \cite{ramage2010characterizing}, revealing the implicit knowledge present in news streams using a multilayer clustering system to support similar topic exploration \cite{pons2007topic}, and using fuzzy latent semantic analysis (FLSA) to eliminate topic redundancies in a medical corpora \cite{karami2018fuzzy}. 

Previous research in constructing knowledge graphs, text mining, and topic retrieval have all used a combination natural language processing techniques and classical machine learning algorithms. Our research focuses on using language models to explore their capabilities in generating topic hierarchies and supporting the process of topic scoping.

\subsection{LLMs on Various Knowledge-Based Tasks}
While many previous approachs leverage classical machine learning and natural language processing techniques to extract information from existing corpora, current research has also focused on leveraging LLMs to support various knowledge-based tasks. Wang et. al. has explored how LLMs can support finding conceptual relations between topics and connecting tangible scenes and experiences with abstract words \cite{wang2023popblends}. Additionally, current work has also been done to support the process of using LLMs to support the clarification of an abstract concept into a semantically-related object \cite{liao2024text}, in sensemaking for complex topics by leveraging LLMs to support multilevel abstractions \cite{suh2023sensecape}, and in retrieval-based knowledge tasks \cite{wu2024stark}. These research areas demonstrate the integration of LLMs to support various knowledge-based tasks, leveraging the ability of LLMs to generate diverse and creative connections between abstract and concrete topics. The topic of topic scoping is related to these works as it uses an LLM to support the systematic retrieval of information from the model itself.

\section{Methodology}
We explore the capabilities of LLMs to incrementally generate topic hierarchies through three different prompting strategies. We generate subtopics for up to 5 different levels of specificity, using a subset of Wikipedia's \href{https://en.wikipedia.org/wiki/Category:Computer_science}{``Category of Computer Science''} page as a test suite of topics. We used human annotators to evaluate the appropriateness of the generated subtopics for each of the prompting strategies.

\subsection{5-Level Topic Classification System}
To standardize the categorization of generated subtopics, we created a 5-level topic hierarchy to classify the level of specificity for a generated topic. Table \ref{tab:5-Levels} illustrates the 5-Level Topic Hierarchy with corresponding descriptions and examples. The table shows that Level 1 is the broadest level and contains topics related to broad domains of study, like Computer Science. The next level, Level 2, contains more specific subtopics that explore general concepts within Computer Science, like Data Structures. Each level's specificity incrementally increases with Level 5 topics being the most specific and focused on specific implementations--like Dijkstra's algorithm as a specific implementation of Shortest Path Algorithms (Level 4 topic). We choose to set the depth of the table to be 5 because preliminary findings have shown that users struggled the most with manually brainstorming topics at Level 5. 

\begin{table}[ht]
    \centering
    \setlength{\arrayrulewidth}{0.8pt} 
    \begin{tabularx}{\columnwidth}{|p{0.14\columnwidth}|p{0.46\columnwidth}|p{0.225\columnwidth}|}
        \hline
        \textbf{Level} & \textbf{Definition} & \textbf{Example in Computer Science} \\ 
        \hline\hline
        Level 1 & Topics related to domains areas of study & Computer Science \\
        \hline
        Level 2 & Subtopics that explore general topics & Data Structures  \\ 
        \hline
        Level 3 & Subtopics that are general concepts & Algorithms  \\
        \hline
        Level 4 & Subtopics exploring different use cases of general concepts & Shortest Path Algorithms  \\
        \hline
        Level 5 & Subtopics that focus on specific implementations & Dijkstra's algorithm \\
        \hline
    \end{tabularx}
    \caption{The corresponding level descriptions in the 5-Level topic classification system.}
    \label{tab:5-Levels}
\end{table}

\begin{table*}[ht]
    \centering
    \setlength{\arrayrulewidth}{0.8pt} 
    \begin{tabularx}{\textwidth}{|p{0.18\textwidth}|p{0.305\textwidth}|p{0.45\textwidth}|}
        \hline
        \textbf{Prompting Strategy} & \textbf{Base Prompt} & \textbf{Level 4 Sample Prompt} \\ 
        \hline\hline
        Current Topic & ``List 5 subtopics of \{curr\_topic\}'' & ``List 5 subtopics of \textit{shortest path algorithms}.'' \\
        \hline
        Root + Current Topic & ``In \{level\_1\}, list 5 subtopics of \{curr\_topic\}'' & ``In \textit{computer science}, list 5 subtopics of \textit{shortest path algorithms}.''\\ 
        \hline
        Full Path + Current Topic & ``In \{level\_1, ..., level\_n-1\}, list 5 subtopics of \{curr\_topic\}'' & ``In \textit{computer science, data structures, and graph algorithms}, list 5 subtopics of \textit{Shortest path algorithms}.'' \\
        \hline
    \end{tabularx}
    \caption{Three different prompting strategies that were used to elicit topic hierarchies from LLMs. The Table illustrates the name of the prompting strategy, the base prompt, and an example prompt using a Level 4 topic of shortest path algorithms.}
    \label{tab:5-Levels-propmting}
\end{table*}

\subsection{Wikipedia's Topic Hierarchy}

After defining the 5-Level classification system, we used Wikipedia's \href{https://en.wikipedia.org/wiki/Category:Computer_science}{``Category of Computer Science''} page as a reference to create test suite of topics to standardize the evaluation process for the different prompting strategies. Wikipedia's ``Category of Computer Science'' page is structured as a nested series of expandable subtopic lists where users are able to incrementally traverse through the different levels. To address any holes in Wikipedia's ``Category of Computer Science'' page, we also referenced online computer science syllabi to supplement any missing pieces of information.

To generate the test suite, we focused on topics in Computer Science (Level 1) as proof of concept. We choose to focus Data Structures, Artificial Intelligence, Databases, and Operating Systems as the four main Level 2 areas. We chose to study these areas because these are common courses in a computer science curriculum. Because we were interested in improving generations at Level 5 topic area, we choose to a total of 20 different subtopics. Figure \ref{fig:sample_category} illustrates the complete test suite that we used to evaluate each prompting strategy. In total we tested 29 different topics in Computer Science. For each topic, we had an LLM generate 5 different subtopics because having multiple options to review is an important step of divergent brainstorming. As a result, a total of 145 generated subtopics were evaluated for each prompting strategy.

\subsection{Prompting Strategies}

We tested three different prompting techniques to help incrementally elicit topic hierarchies following the 5-Levels Topic Classification System. We specifically used OpenAI's GPT-4 API \cite{gpt4} as the LLM for this task and tested each prompting strategy on the Wikipedia test suite of Computer Science concepts. For all prompting strategies, we explicitly asked GPT-4 to generate 5 subtopics. The three prompting strategies are illustrated in Table \ref{tab:5-Levels}. The Current Topic prompting strategy only contains the current topic when generating subtopics. The Root + Current Topic prompting strategy contains both the Level 1 topic, or root, and the current topic. Finally, the Full Path + Current Topic prompting strategy includes the entire chain of parent topics, leading up and including the current topic in the prompt.

\subsection{Evaluation Criteria}
The first author and an independent expert were tasked with annotating the generated subtopics for each prompting strategy. Both annotators were experts in computer science. We provided the evaluation rubric below to each annotator, along with detailed directions with how to annotate the generated subtopics. Annotations were done separately. Each annotator labeled 145 generated subtopics for each of the 3 different prompting strategies. We paid \$16 an hour for 4 hours of work.  

The evaluation criteria given to annotators cover issues of repetitiveness, specificity, and relatedness to the inputted topic:
\begin{enumerate}[noitemsep]
    \item \textbf{Repetitive}: the generated subtopic repeats the same input topic; 
    \item \textbf{Too specific}: the generated subtopic is too specific for the desired level;
    \item \textbf{Too general}: the generated subtopic is broader than the desired level;
    \item \textbf{Tangential}: the generated subtopic is at the correct level of specificity but is not directly related to the input topic;
    \item \textbf{Unrelated}: the generated subtopic is unrelated to the root level topic.
\end{enumerate}

\section{Results} 
\begin{figure}
    \centering
    \includegraphics[width=\columnwidth]{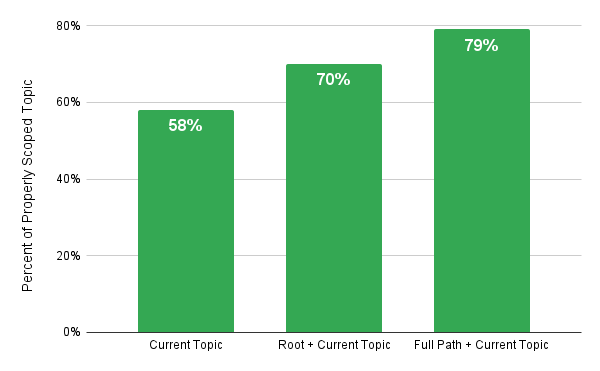}
    \caption{Bar chart demonstrating the percentage of properly scoped subtopics across all levels for each prompting strategy.}
    \label{fig:percent-good}
\end{figure}

The two annotators had a substantial inter-rater agreement on their assessment over three strategies, an average Cohen-Kappa of 0.61 across all annotation assignments. As a result we averaged the accuracy across the two annotators because of the high agreement. The Full Path + Current Topic yielded the highest average accuracy of \textbf{77\%}. Followed by Root + Current Topic with an accuracy of 70\%, and then Current Topic with an accuracy of 58\%. These results are demonstrated in Figure \ref{fig:percent-good}. These results show that by including the full path of parent topics helped GPT-4 generate more concrete and specific subtopics. Including Root + Current Topic helped more than providing no additional information to the base prompt, Current Topic. This demonstrates that by providing more context in the prompt helps GPT-4 with generating specifically scoped and concrete subtopics.

\begin{table*}[t]
\resizebox{\textwidth}{!}{%
\begin{tabular}{l|ccccc|}
\cline{2-6}
                                                         & \multicolumn{5}{c|}{\textit{Improperly Scoped Topics}}                                                                                                        \\ \cline{2-6} 
                                                         & \multicolumn{1}{c|}{\textbf{Too General}} & \multicolumn{1}{c|}{Too specific} & \multicolumn{1}{c|}{Unrelated} & \multicolumn{1}{c|}{Tangential} & Repetitive \\ \hline
\multicolumn{1}{|l|}{\textbf{Current Topic}}             & \multicolumn{1}{c|}{\textbf{27\%}}        & \multicolumn{1}{c|}{8\%}          & \multicolumn{1}{c|}{4\%}       & \multicolumn{1}{c|}{2\%}        & 1\%        \\ \hline
\multicolumn{1}{|l|}{\textbf{Root + Current Topic}}      & \multicolumn{1}{c|}{\textbf{14\%}}        & \multicolumn{1}{c|}{9\%}          & \multicolumn{1}{c|}{3\%}       & \multicolumn{1}{c|}{3\%}        & 0\%        \\ \hline
\multicolumn{1}{|l|}{\textbf{Full Path + Current Topic}} & \multicolumn{1}{c|}{\textbf{9\%}}         & \multicolumn{1}{c|}{9\%}          & \multicolumn{1}{c|}{0\%}       & \multicolumn{1}{c|}{2\%}        & 0\%        \\ \hline
\end{tabular}%
}
\caption{Distribution of error category across each prompting strategy for all generated subtopics.}
\label{tab:error-breakdown}
\end{table*}

We performed an analysis on the generated subtopics and found that the biggest problem was Too General errors, followed by Too Specific errors. Rarely were generated subtopics Tangential or Repetitive. For the Current Topic prompting strategy, 27\% of the errors were due to the generated subtopics being Too General, while the Root + Current Topic and the Full Path + Current Topic yielded 14\% and 10\% improperly scoped topics due to Too General errors. While Too General errors were seen as the largest source of error across all three prompting strategies, Too Specific errors also accounted for the second largest error category, averaging about 9\% of the errors for all prompting approaches. While prompting strategy reduced many of the Too General errors, the number of Too Specific errors remained consistent across all prompting strategies. Table \ref{tab:error-breakdown} contains the distribution  of error type across all 3 different prompting strategies. 

We also analyzed the specific levels that each type of error was occurring, as seen in Table \ref{tab:errors-by-level}. We found that that most errors occurred at Level 5.  Too General errors were the most frequent error type at Level 5. This demonstrates that LLMs struggle to elicit specific information at that depth. Because LLMs operate on generating the next most probably word, it is possible that the more specific the subtopic, the less probable it is. As a result, LLMs will revert back to general topics when a certain level of specificity is reached because the generating a general topic might have a higher likelihood than generating a specific subtopic. 

The Full Path + Current Topic prompting strategy reduced the errors in Level 5. But errors generated in Level 2, Level 3, and Level 4 were not reduced by the Full Path + Current Topic prompting strategy and remained largely similar across all prompting techniques. At Level 2 and Level 3, Too Specific was one of the most common errors generated. The reason that Too Specific errors occurred at the broader levels of specificity is probably due to LLMs struggling to incrementally generate subtopics. Specific examples of these errors are listed in the following sections.

\begin{table}[]
\resizebox{\columnwidth}{!}{%
\begin{tabular}{l|c|c|c|c|}
\cline{2-5}
                                                         & \multicolumn{1}{l|}{\textbf{Level 2}} & \multicolumn{1}{l|}{\textbf{Level 3}} & \multicolumn{1}{l|}{\textbf{Level 4}} & \multicolumn{1}{l|}{\textbf{Level 5}} \\ \hline
\multicolumn{1}{|l|}{\textbf{Current Topic}}             & 1\%                                   & 4\%                                   & 2\%                                   & 34\%                                  \\ \hline
\multicolumn{1}{|l|}{\textbf{Root + Current Topic}}      & 1\%                                   & 3\%                                   & 4\%                                   & 21\%                                  \\ \hline
\multicolumn{1}{|l|}{\textbf{Full Path + Current Topic}} & 1\%                                   & 3\%                                   & 5\%                                   & 12\%                                  \\ \hline
\end{tabular}%
}
\caption{Distributions of errors across each level for each prompting strategy.}
\label{tab:errors-by-level}
\end{table}

\subsection{Too General Error Examples} 
An example of a generated Level 5 subtopic being too General is for the topic of ``Minimum Spanning Trees'' and the generated subtopic ``Randomized Algorithms'' as a subtopic. ``Minimum Spanning Trees'' are graphs that connect all vertices with the minimum possible total edge weight while ``Randomized Algorithms'' is any algorithm that uses any degree of randomness in its logic. ``Randomized Algorithms'' are too general because they are not a specific implementation of minimum spanning trees, instead they are just a subset of types of algorithms. An appropriate subtopic of ``Minimum Spanning Trees'' would be ``Kruskal's Algorithm'' because that is a specific algorithm that is used for the purpose of finding minimum spanning trees. 

Less frequently, Too General errors occur at the Level 3 area due to overlaps between the topic and the generated subtopic. For example, ``Robotics'' was generated as a subtopic of ``Artificial Intelligence,'' a field that involves creating machines to emulate human intelligence. While there is an intersection between robots and AI techniques, the broad domain of robotics does not require robots to emulate human intelligence in the same way subtopics in ``Artificial Intelligence'' do. Wikipedia classifies Robotics as a subtopic under Computer Engineering and not a subtopic under Artificial Intelligence. As a result, the generated subtopic of ``Robotics'' under ``Artificial Intelligence'' is too specific. 

\subsection{Too Specific Error Examples in Level 3} 
The Level 2 topic of ``Artificial Intelligence'' and the generated subtopic of ``Neural Networks'' is another example of the generated subtopics being Too Specific at the Level 3 area. Neural Networks are a specific structure that are used in machine learning and should be classified as a Level 4 topic which covers topics that explore different examples. To correct this example, the sequence of topics should go from ``Artificial Intelligence'' to ``Machine Learning'' and then to ``Neural Networks.'' This sequence of topics first covers the general field of how machines can emulate human behaviors, before getting more specific into how machines can learn like humans, and finally how to use a technique to model how the human brain learns in computers.

\section{Discussion}

Building off previous research that use classical machine learning methods to extract topic hierarchies from unstructured data, we explored the effectiveness of LLMs in eliciting topic hierarchies and supporting the topic scoping processing: moving from abstract concepts to concrete examples. While LLMs are effective in generating a diverse set of subtopics, they still need assistance in structuring the topics into a topic hierarchy. We developed a 5-levels topic classification structure and used Wikipedia's ``Category of Computer Science'' page as a test suite of topics for each of the 5-levels. We explored 3 different prompting techniques to elicit topic hierarchies from LLMs and found that by including the entire sequence of parent topics helped reduce issues of improperly scoped topics. 

\subsection{Applications}
There are many areas where dynamically generating topic hierarchies are beneficial. For example: educators can leverage the process of topic scoping to assist in curriculum development, content creators can use dynamically generated topic hierarchies to explore specific niches within larger themes, and product managers can use topic scoping to break down abstract goals into smaller, more concrete subtasks.

This work can be applied to the field of education to help educators with curriculum development. Traditionally, the curriculum design process requires the educator to conduct extensive planning and research based on the previous curricula and the specific needs of the students \cite{curriculumDesign}. Through topic scoping, educators can better design a more audience-centric curriculum that fits their specific student demographics. The process of topic scoping and recreating knowledge hierarchies could help educators decide which topics to cover over the course of the semester that align with the grade level of their class. 

Additionally, topic scoping can be used for content creators when brainstorming the types of topics they should cover. Topic scoping can help creators narrow down broad interests like reading into more specific ideas. For examples, Max is a content creator on Tiktok for book and reading content. He wants to explore what videos he should film for the upcoming month, he can use topic scoping to find different types of books to explore. Books can be broken down by genres like Romance and Fiction, these genres can then be explored by setting, time period, and author. Max wants to explore Contemporary Fiction, from there he can further narrow down this topic into Translated, Contemporary Fiction, and even further into Japanese-Translated, Contemporary Fiction. Here Max is ready to start brainstorming his list of top 5 Japanese-Translated, Contemporary Fiction books to share with his Tiktok followers. Just like that, topic scoping helped Max find a niche within a larger reading community. 

More broadly, the process of breaking an abstract goal into a series of concrete tasks and sub-goals can be applied to project and product managers as they track the development of a project by the smaller subgoals towards a more abstract goal. For example, Jenna is a product manager that is in charge of developing a new feature for a e-commerce website to improve traffic. The broad goal of feature development to  improve traffic is abstract, so Jenna might break down that goal into smaller sub-goals like understanding current user traffic data on the site and even more specific goals like doing user studies and A/B tests. The applications of topic scoping are demonstrated in the process of iteratively, narrowing down a goal into specific and concrete tasks. By creating a topic hierarchy, Jenna is able to track the overall progress of the project and work in parallel with her teammates by each tackling one sub-goal category. 

\subsection{Limitations}
While including the all parent topic yielded 78\% of properly scoped subtopics, the main issue that caused improper subtopic generations were due to scope. Generated subtopics errors were mostly likely to be either Too General or Too Specific, demonstrating that more research can be done to improve the scoping issue. One approach is to formalize more fine-grained definitions for each level in the 5-levels classification system and including the definitions in in the prompt to improve Level 5 generations. Additionally, more work can be done to explore specific prompting techniques to reduce the amount of errors generated at Levels 2 and 3 since the current prompting strategies don't reduce the errors at the broader levels.

While there might be concerns around GPT-4 already being trained on Wikipedia's ``Category of Computer Science'' topics, our research isn't focused on novel topic generation instead we are exploring whether LLMs can incrementally generate topics at an increasing level of specificity. As a result, having a model trained on this data should not significantly impact performance. Additionally, our findings show that even if the model was trained on Wikipedia's ``Category of Computer Science'' topics, GPT-4 still struggles to generate specific and concrete subtopics at Level 5. This demonstrates that despite possibly training on this data, the model still struggles with replicating the hierarchical structure. 

Finally, this work focuses on generating subtopics in Computer Science as a proof of concept. Future work can explore how these strategies can be generalized to other domains like Biology, Chemistry, and Physics. Additionally, an interactive, user-driven interface can be developed to support users in the process of topic scoping.

\section{Conclusion}
We found that it is possible to generate topic hierarchies from LLMs by incrementally generating subtopics at increasing levels of specificity. This finding allows for future work to be done on dynamically generating, user-directed topic hierarchies to support a range of tasks like curriculum development for educators, content creation for creators, and product management tools.



\bibliographystyle{ACM-Reference-Format}
\bibliography{main}

\end{document}